\documentclass{PoS}
\usepackage{epsfig}
\usepackage{latexsym}
\usepackage{bm}

\title{Confinement and Chiral Symmetry, a Lattice QCD test of AdS/QCD }

\ShortTitle{Confinement and Chiral Symmetry, a Lattice QCD test of AdS/QCD }

\author{\speaker{D.~K.~Sinclair}%
\thanks{This research was supported in part by US Department of Energy
contract DE-AC02-06CH11357, and in part under a Joint Theory Institute
grant.}\\
        HEP Division, Argonne National Laboratory, 9700 South Cass Avenue,
        Argonne, IL 60439, USA\\
        E-mail: \email{dks@hep.anl.gov}}

\abstract{We use lattice QCD simulations to test some of the predictions of
proposed AdS/QCD (holographic) duals for QCD. In particular, these duals
predict that the scale of chiral symmetry breaking ($\chi$SB) can be varied
independently from that of confinement, with the proviso that the scale of
$\chi$SB cannot be longer than that of confinement. We simulate lattice QCD
with 2 quarks in the fundamental representation of colour and with additional
4-fermion interactions (suggested by AdS/QCD), at finite temperatures. For
sufficiently strong 4-fermion interactions, the deconfinement and $\chi$SB
transitions occur at different temperatures, the separation depending on the
4-fermion coupling. This confirms that the scales of confinement and $\chi$SB
are, in general, different.}

\FullConference{The XXVI International Symposium on Lattice Field Theory \\
		 July 14 - 19, 2008\\
		 Williamsburg, Virginia, USA}

\begin{document}

\section{Introduction}

With standard actions, simulations of the finite temperature phase structure
of lattice QCD with fundamental quarks indicate that the deconfinement 
transition and the chiral-symmetry restoration transition are coincident
\cite{Polonyi:1984zt}.
For quarks in other representations of $SU(3)_{colour}$, deconfinement occurs
at a lower temperature than chiral-symmetry restoration 
\cite{Kogut:1984sb,Engels:2005te}.

On the other, hand suggested holographic duals of QCD with fundamental quarks,
inspired by AdS/CFT duality \cite{Karch:2002sh,Sakai:2004cn,Sakai:2005yt}, 
predict that the scales of confinement and of
chiral symmetry breaking ($\chi$SB) can be varied independently
\cite{Antonyan:2006vw,Aharony:2006da}. 
However, since 
confinement produces $\chi$SB, the length scale associated with $\chi$SB
cannot be greater than that associated with confinement 
\cite{Banks:1979yr,Leutwyler:1992yt}.

These proposed holographic (string/gravity) duals of QCD suggest that the
scales of confinement and $\chi$SB can be decoupled by the addition of 
(non-local) 4-fermion interactions to QCD. Adding attractive 4-fermion 
interactions binds the quarks and antiquarks more tightly so that $q\bar{q}$
pairs can condense and spontaneously break chiral symmetry at shorter distances.

We consider lattice QCD with 2-flavours of staggered quarks and {\it local}
4-fermion interactions of the Gross-Neveu \cite{Gross:1974jv}%
/Nambu-Jona-Lasinio \cite{Nambu:1961tp,Nambu:1961fr} type. 
We work
at finite temperature and use the deconfinement and chiral-symmetry restoration
temperatures as our measure of the scales of confinement and $\chi$SB. These
are measured as functions of the 4-fermion coupling. These simulations are
described in reference~\cite{Sinclair:2008du}

\section{The lattice action}

The lattice quark action is the standard $\chi$QCD action
\cite{Kogut:1998rg}, which is the
traditional staggered quark action augmented by a chiral 4-fermion term.
Expressed in terms of the auxilliary fields $\sigma$ and $\pi$ to render it
quadratic in the fermion fields it is
\begin{equation}
S_f = \sum_{f=1}^{N_f/4}\sum_s\bar{\chi}_f\left[\not\!\! D+m+
        \frac{1}{16}\sum_i(\sigma_i+i\epsilon\pi_i)\right]\chi_f 
    + \sum_{\tilde{s}}\frac{1}{8}N_f\gamma(\sigma^2+\pi^2)
\end{equation}
$\epsilon=(-1)^{x+y+z+t}$ and $\gamma$ is the inverse 4-fermion coupling.
This preserves the exact $U(1)$ axial flavour symmetry of staggered fermions.

Simulations are performed using the exact RHMC algorithm \cite{Clark:2006wp}
to tune to 2 flavours.
The deconfinement transition is determined as the position of the rapid 
increase in the Wilson Line (Polyakov Loop). The chiral-symmetry restoration
phase transition occurs where the chiral-condensate $\bar{\psi}\psi$ vanishes
($m=0$).

\section{Simulations and Results}

If we turn off the QCD interactions, we are left with a 4-fermion model which
has a bulk transition at $\gamma=\gamma_c \approx 1.7$. At finite temperature,
this will be shifted to smaller $\gamma$ (stronger coupling).

Our finite temperature simulations are performed on $N_t=4$ lattices. We keep
$\gamma > \gamma_c$ so that at high temperatures -- weak gauge coupling (large 
$\beta=6/g^2$) -- the theory is in the chiral-symmetry restored phase. Previous
simulations at $\gamma=10$ and $\gamma=20$ indicate that the deconfinement
and chiral-symmetry restoring transitions are coincident \cite{Kogut:2002rw}. 
Hence $\gamma \ge 10$
represents a weak 4-fermion coupling. We simulate at $\gamma=2.5$, a strong
4-fermion coupling and $\gamma=5$, and intermediate coupling. $N_f=2$ and $m=0$.
At $\gamma=2.5$ our lattice sizes are $16^3 \times 4$, $24^3 \times 4$ and
$32^3 \times 4$. At $\gamma=5$ our lattice sizes are $12^2 \times 24 \times 4$,
$24^3 \times 4$ and $32^3 \times 4$. Typical run lengths for each parameter set
are 50,000 or 100,000 trajectories.

\begin{figure}[htb]
\epsfxsize=3in
\epsffile{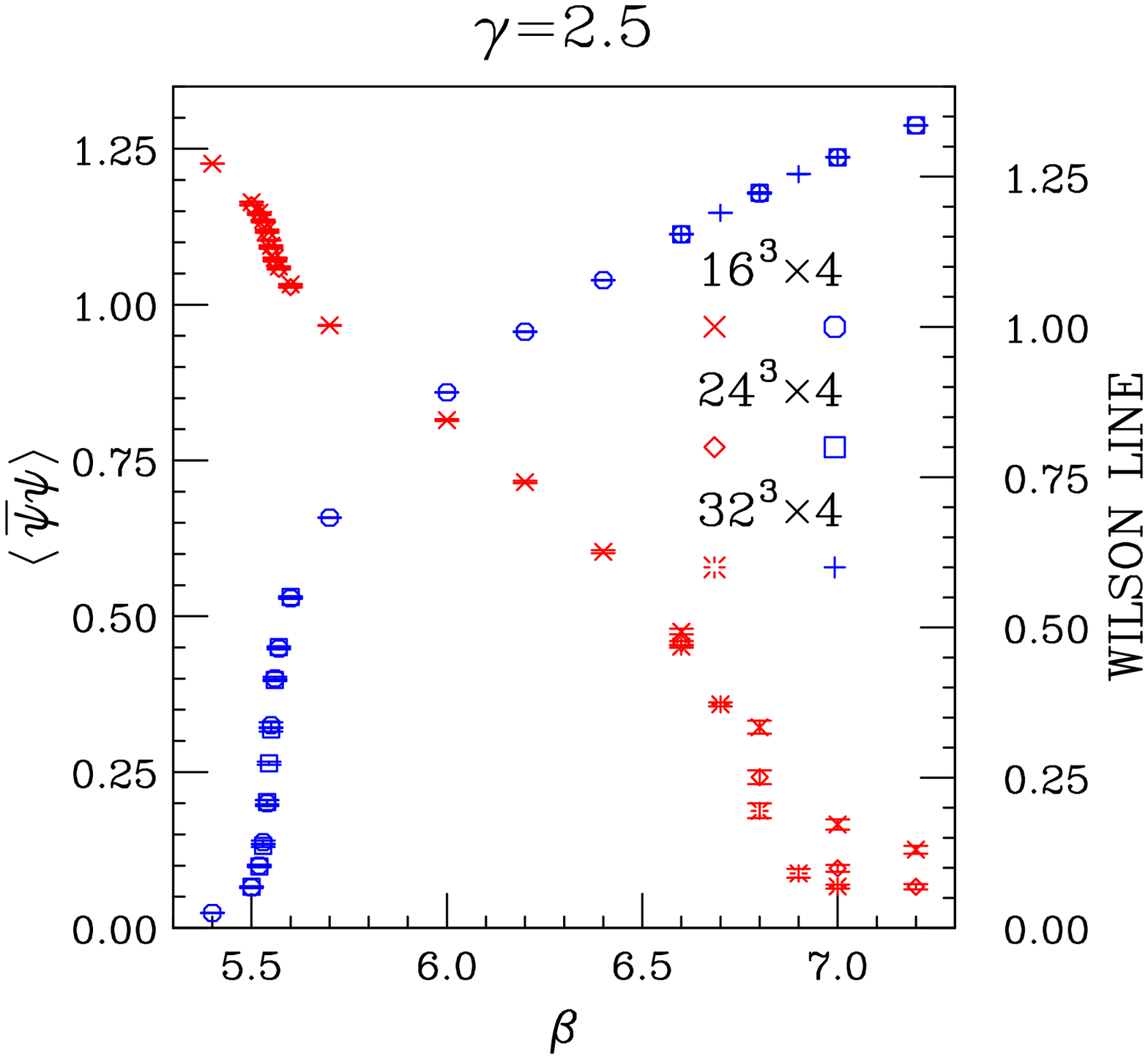}
\epsfxsize=3in
\epsffile{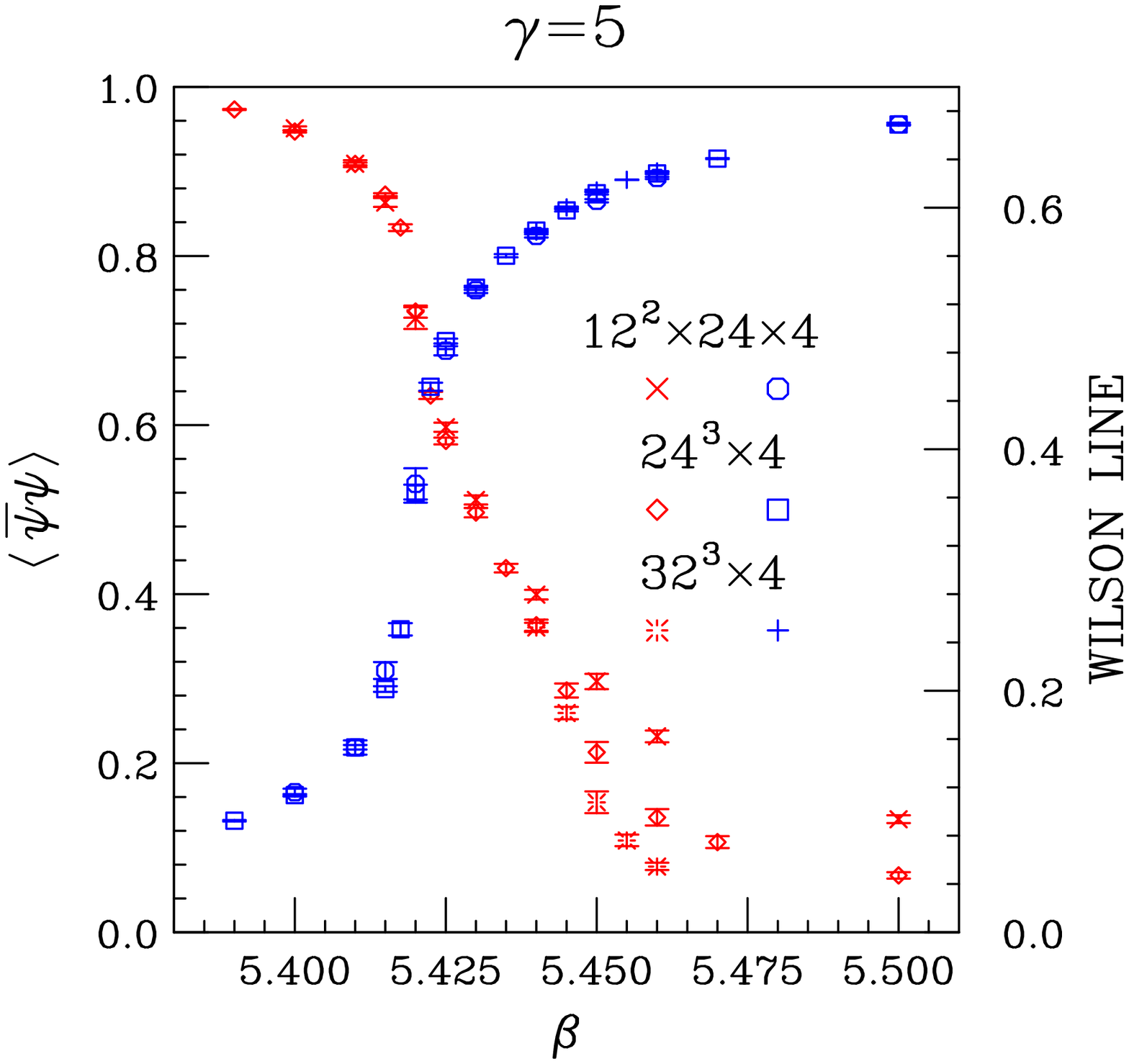}
\caption{Wilson line and chiral condensate as functions of $\beta$ for
(a) $\gamma=2.5$ and (b) $\gamma=5$ in lattice units.}
\label{fig:wil-psi}
\end{figure}

Figure~\ref{fig:wil-psi} shows the Wilson Line and chiral condensate as
functions of $\beta$ for the two $\gamma$s considered. At $\gamma=2.5$ the
deconfinement transition and the chiral-symmetry restoration transition are
well separated. At $\gamma=5$ the two transitions are close, but still clearly
separate. It is clear that the finite size effects are small near the
deconfinement transition. As expected, the finite size effects are
considerable close to the chiral-symmetry restoration transition. These graphs
include points (and for $\gamma=2.5$ a lattice size) in addition to those in
our earlier publication \cite{Sinclair:2008du}.

\begin{figure}[htb]
\epsfxsize=3in
\epsffile{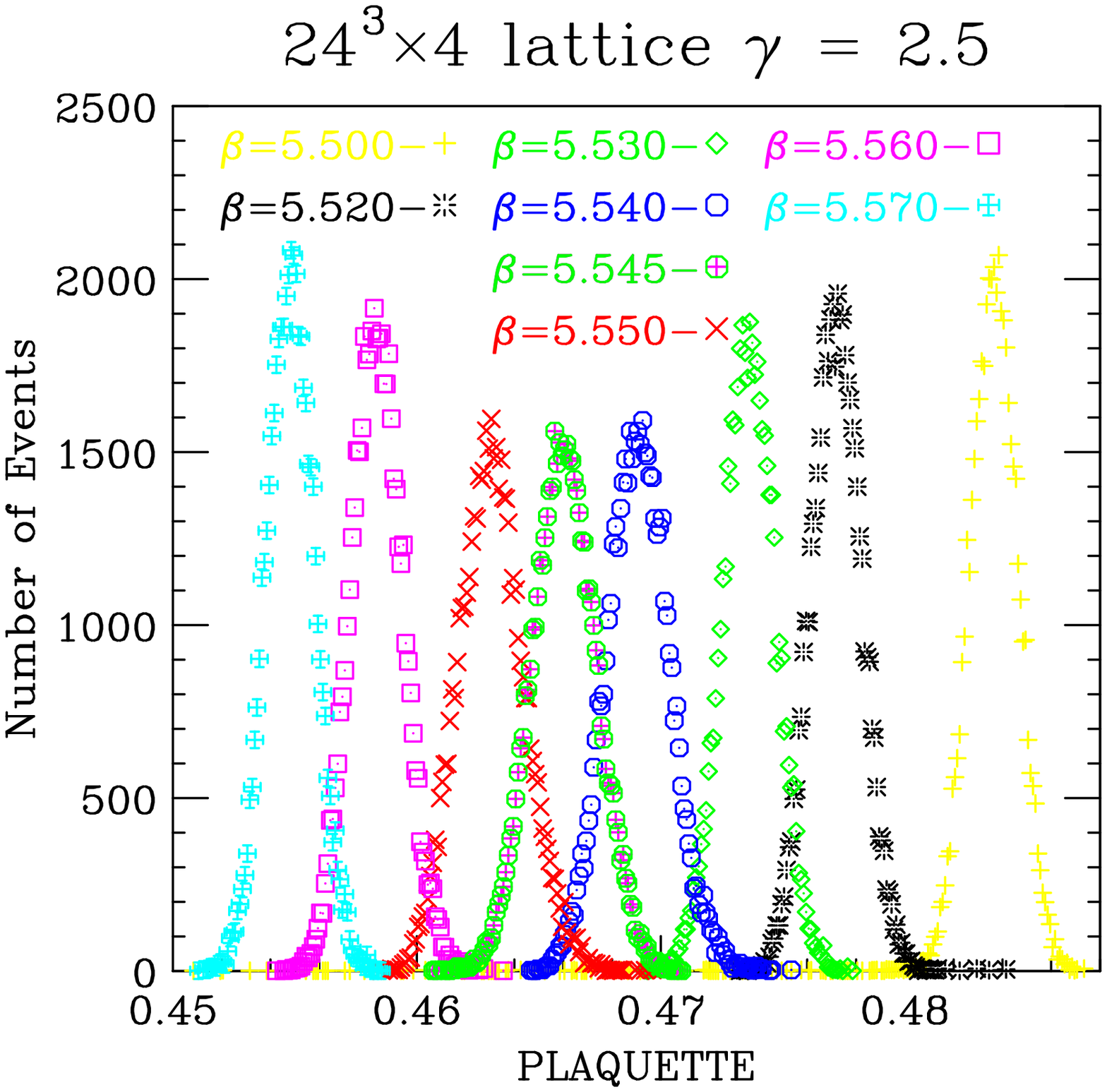}
\epsfxsize=3in
\epsffile{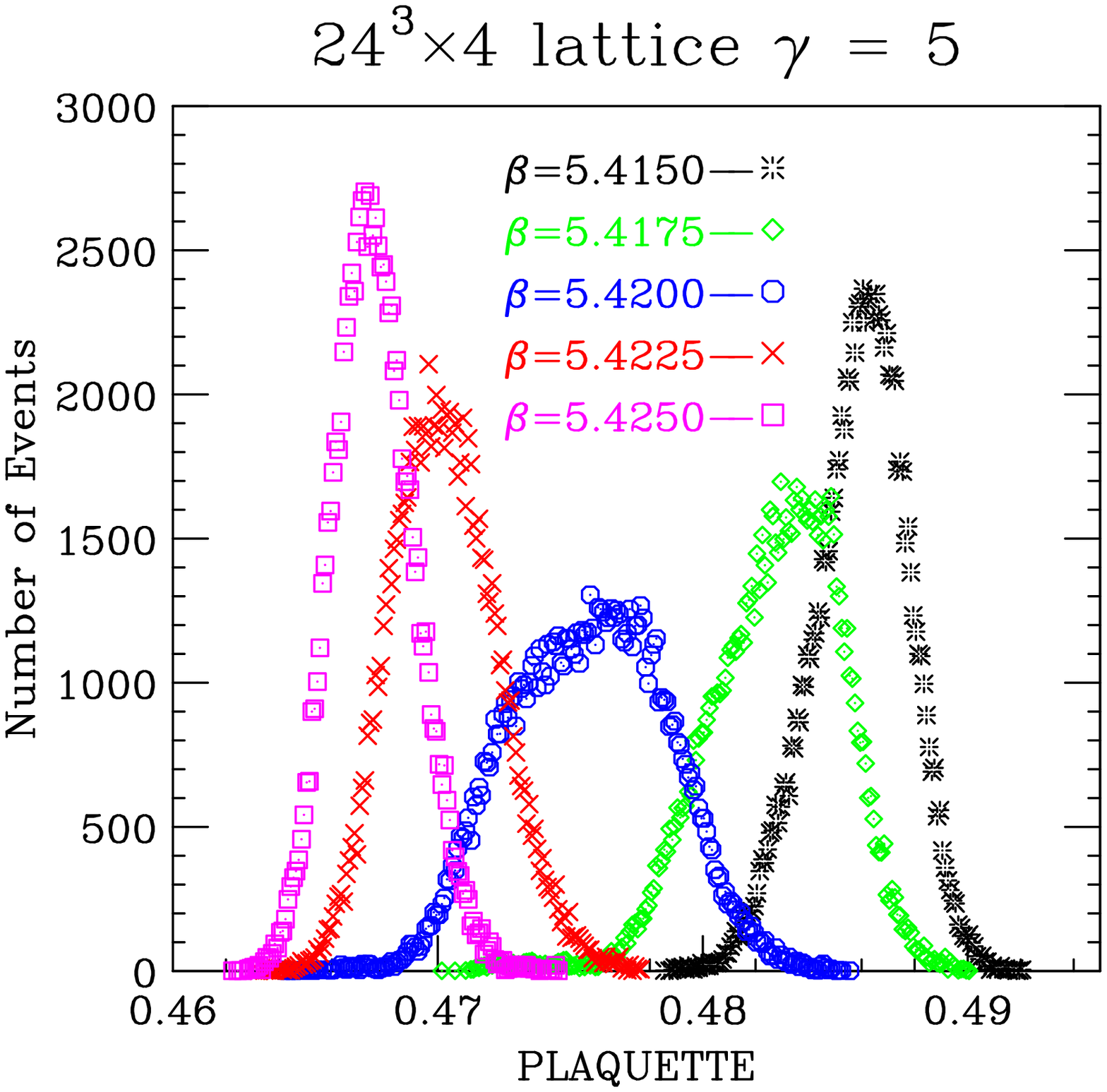}             
\caption{a) Plaquette distributions close to the deconfinement transition on
a $24^3 \times 4$ lattice at $\gamma=2.5$. (Since we have twice the statistics
at $\beta=5.545$ as at the other $\beta$s, we have divided these points by 2).
b) Same, but for $\gamma=5$.}
\label{fig:plaq-hist-c}
\end{figure}

\begin{figure}[htb]
\epsfxsize=3in
\centerline{\epsffile{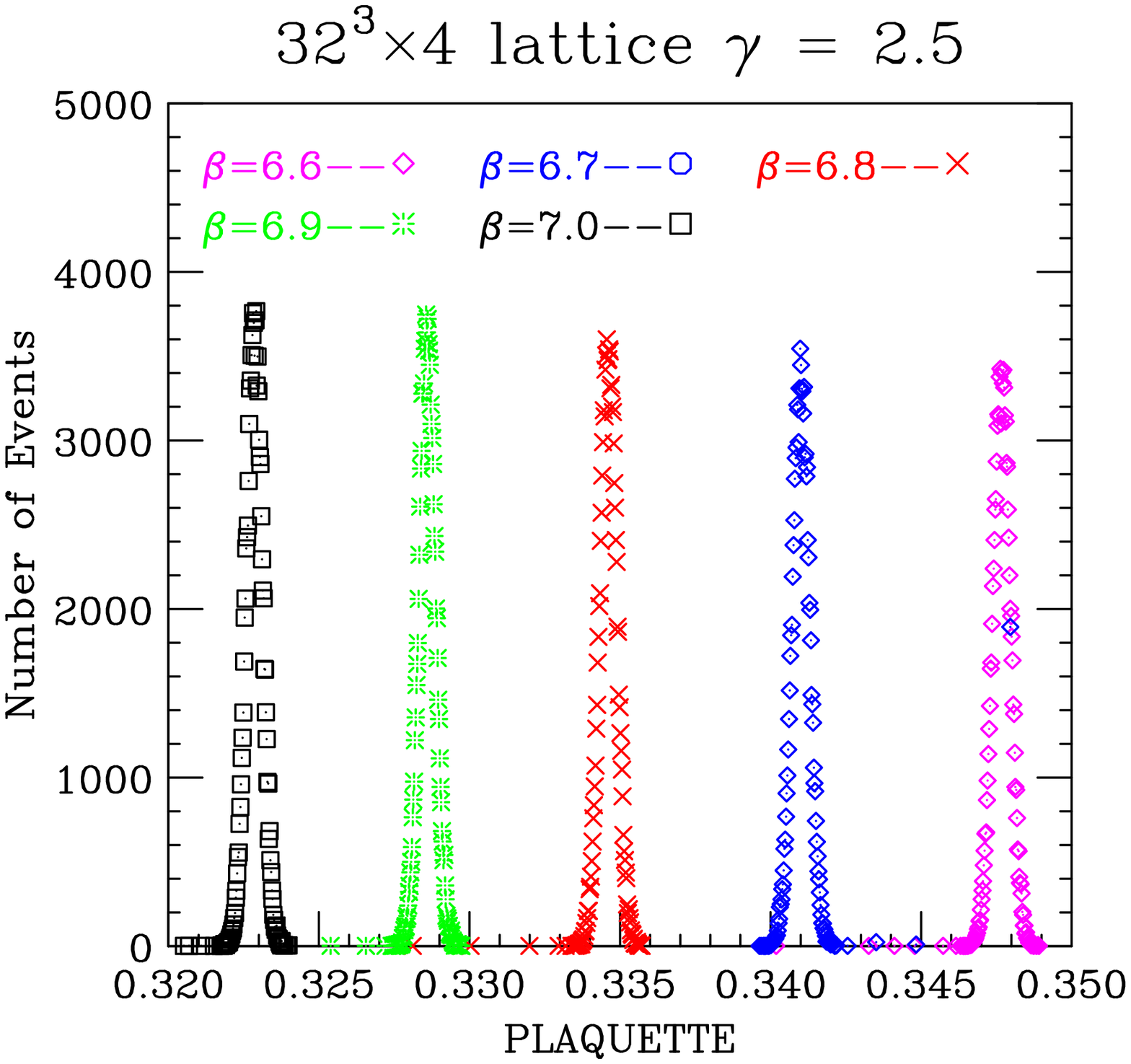}}
\epsfxsize=3in
\centerline{\epsffile{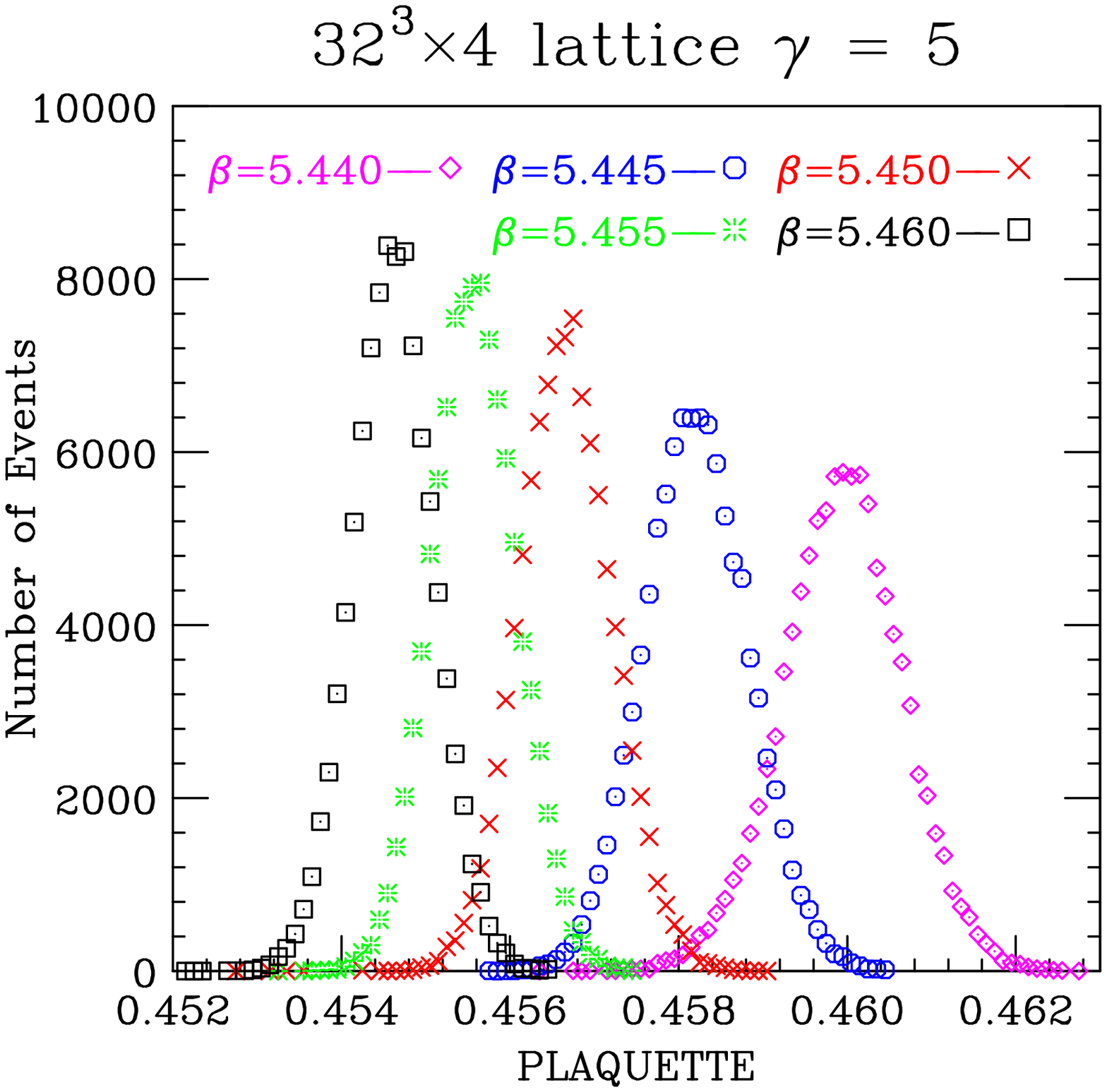}}
\caption{Plaquette distributions close to the chiral-symmetry restoration
transition on a $32^3 \times 4$ lattice (a) at $\gamma=2.5$, (b) at $\gamma=5$.}
\label{fig:plaq-hist-chi}
\end{figure}

At $\gamma=2.5$ we estimate the position of the deconfinement transition from
the peak in the Wilson Line susceptibility, using Ferrenberg-Swendsen 
reweighting \cite{Ferrenberg:1988yz}
from $\beta=5.545$. This yields $\beta_d=5.547(3)$. This is
possible because the distributions of plaquette values from the simulated
$\beta$s in the neighbourhood of this transition overlap (see
figure~\ref{fig:plaq-hist-c}a). The $\beta$s for the simulations near the
chiral transition are not close enough for such an estimate (see
figure~\ref{fig:plaq-hist-chi}a). Our (more subjective) estimate for this chiral
transition is $\beta_\chi=6.85(5)$.

For $\gamma=5$, we estimate the position of the deconfinement transition to
be $\beta_d=5.420(4)$. For the chiral transition we estimate 
$\beta_\chi=5.450(5)$.

Figures~\ref{fig:plaq-hist-c},\ref{fig:plaq-hist-chi} show the plaquette 
distributions near the deconfinement and chiral transitions.

The deconfinement $\beta$, $\beta_d$, is restricted to the range
$\beta_d(\gamma=\infty) \le \beta_d \le \beta_d(\gamma=\gamma_c)$ with 
$\beta_d(\gamma=\gamma_c) \le \beta_d(quenched)$. Hence 
$5.25 \lesssim \beta_d \lesssim 5.6925$. [The lower bound (zero 4-fermion
coupling) is from \cite{Bernard:1999fv,D'Elia:2005bv} while the upper limit
(quenched QCD) is from \cite{Brown:1988qe,Bacilieri:1988yq}.]
The chiral-symmetry restoring $\beta$, $\beta_\chi$, is in the range
$\beta_d \le \beta_\chi \le \infty$, where the lower bound is achieved for
small 4-fermion coupling, and the upper for $\gamma=\gamma_c$. 
Table~\ref{tab:} summarises these results. (Note that we have not included
error bars on the results of others.) 

\begin{table}[htb]
\centerline{
\begin{tabular}{l@{\hspace{1in}}l@{\hspace{1in}}l}
\hline
$\gamma$               &$\beta_d$        &$\beta_\chi$         \\
\hline
0.0                    &5.6925           &  ---                \\
$\gamma_c \approx 1.7$ &   ?             &$\infty$             \\
2.5                    &5.547(3)         &6.85(5)              \\
5.0                    &5.420(4)         &5.450(5)             \\
10.                    &5.327(2)         &5.327(2)             \\
20.                    &5.289(1)         &5.289(1)             \\
$\infty$               &5.25             &5.25                 \\
\hline
\end{tabular}
}
\caption{Deconfinement and chiral-symmetry restoration $\beta$s as functions
of $\gamma$.}
\label{tab:}
\end{table}

\section{Discussion and Conclusions}

\begin{itemize}
\item
Adding extra 4-fermion interactions with sufficient strength to the lattice QCD
action separates the deconfinement and chiral-symmetry restoration transitions
at finite temperatures. We are able to change the separation of the scales
of confinement and $\chi$SB by varying the 4-fermion coupling, as predicted
from proposed holographic duals of QCD.
\item
$5.25 \lesssim \beta_d \lesssim 5.6925$, while 
$\beta_d \le \beta_\chi \le \infty$.
\item
At $\gamma=2.5$ -- strong 4-fermion coupling -- $T_\chi \sim 10 T_d$.
At $\gamma=5$ -- intermediate 4-fermion coupling -- $T_\chi \sim 1.04 T_d$.
For $\gamma \ge 10$ -- weak coupling -- $T_\chi = T_d$.
\item
We should perform simulations at more $\gamma$ values in the range
$2 \lesssim \gamma \lesssim 7$. Our simulations used $N_t=4$. A more complete 
study should include other $N_t$ values. 
\item
We have qualitative agreement with holographic QCD. We need to make the
comparison more quantitative.
\item
Is the deconfinement transition a phase transition, or merely a crossover?
As the 4-fermion coupling increases to infinity, we expect the theory to 
approach quenched QCD, where deconfinement is a first order transition.
We also see some indication of this transition becoming stronger as the 
4-fermion coupling weakens and the two transitions approach one another.
\item
With a little more work we should be able to determine the universality class
of the second order chiral transition.
\item
The two phase transitions appear to coalesce at a finite (non-zero)
4-fermion coupling ($\gamma$ a little larger than 5).
\item
We used a local irrelevant 4-fermion interaction, which will not survive the
continuum limit. Does the non-local 4-fermion interaction indicated by the
AdS/QCD models survive the continuum limit, i.e. does it define a 
non-perturbatively renormalizable theory? An intermediate model would be
a Yukawa model where the auxilliary fields have full 4-dimensional scalar 
dynamics. Such a model is (or can be) perturbatively renormalizible, but
is believed to be non-perturbatively trivial.
\item
Theories with quarks in higher representations of the colour group, where the
stronger QCD coupling separates the confinement and $\chi$SB scales even without
the extra 4-fermion terms, should be studied. The Bielefeld group have studied
adjoint quarks where the transitions are separated and are required by symmetry
to both be phase transitions \cite{Engels:2005te}. 
Colour sextet quarks are also of some interest,
especially with regard to conformal technicolor models \cite{Shamir:2008pb}. 
\end{itemize}

\section*{Acknowledgements}
These simulations were performed on the Cray XT4, Franklin, at NERSC. Earlier
work used the University of Kentucky HP Superdome. We thank Jeffrey Harvey and
David Kutasov for raising the questions which inspired this research. We also
acknowledge helpful discussions with Cosmas Zachos and Alexander Velytsky.

\end{document}